\documentclass[]{spie}  

\usepackage{amsmath,amsfonts,amssymb}
\usepackage{graphicx}
\usepackage[colorlinks=true, allcolors=blue]{hyperref}
\usepackage{mathabx}

\title{Optimizing MARVEL for the radial velocity follow-up of TESS and PLATO transiting exoplanets}

\author[a]{Cyprien Lanthermann}
\author[a]{Joris De Ridder}
\author[a]{Hugues Sana}
\author[a]{Pierre Royer}
\author[a]{Denis Defrère}
\author[a]{Gert Raskin}
\author[a]{Bart Vandenbussche}
\author[a]{Andrew Tkachenko}
\author[a]{Hans Van Winckel}
\affil[a]{Institute of Astronomy, KU Leuven, Celestijnenlaan200D, 3001, Leuven, Belgium}

\authorinfo{Further author information: (Send correspondence to C.L.)\\C.L.: E-mail: cyprien.lanthermann@kuleuven.be}

\begin{document} 
\maketitle

\begin{abstract}
The space missions TESS and PLATO plan to double the number of 4000 exoplanets already discovered and will measure the size of thousands of exoplanets around the brightest stars in the sky, allowing ground-based radial velocity spectroscopy follow-up to determine the orbit and mass of the detected planets. The new facility we are developing, MARVEL (Raskin et al. this conference~\cite{MARVEL}), will enable the ground-based follow-up of large numbers of exoplanet detections expected from TESS and PLATO, which cannot be carried out only by the current facilities that achieve the necessary radial velocity accuracy of 1~m\,s$^{-1}$\, or less. This paper presents the MARVEL observation strategy and performance analysis based on predicted PLATO transit detection yield simulations. The resulting observation scenario baseline will help in the instrument design choices and demonstrate the effectiveness of MARVEL as a TESS and PLATO science enabling facility.
\end{abstract}


\section{INTRODUCTION}
\label{sec:intro}  

Since the first detection of a planet orbiting a Sun-like star outside our solar system~\cite{Mayor1995}, astronomers have detected more than four thousand exoplanets. These discoveries  were made possible thanks to a variety of  techniques. Two of them in particular have been driving these numbers. Precision photometry allows the detection of exoplanets transiting in front of their host star while  spectroscopy measures the periodic shift of radial velocity (RV) of the host star due to its reflex movement around the center of mass of the joint star-planet system. Both techniques have been applied from the ground and in space.

Nowadays, the effort is shifting from the detection of exoplanets to their characterization, intending to find earth analogs in the habitable zone of their host star, and ultimately to discover the signature of potential bio-markers in their atmosphere.

In this context, future space missions will measure the size of thousands of new exoplanets thanks to the transit detection methods (NASA/TESS, ESA/PLATO), and the space  ESA/ARIEL mission will characterize the atmosphere of  known exoplanets through infrared (IR) spectroscopy. 

While these missions will yield ground-breaking knowledge on a tremendous amount of exoplanets, they will require multi-epoch, ground-based, high-accuracy spectroscopy. Indeed, the mass and density of an exoplanet -- as well as its absolute distance from the star -- can only be computed thanks to the combination of both transit and radial velocity techniques. Furthermore, the characterization of the stellar activity of the host star through visible spectroscopy is required for ARIEL to extract the exoplanet signal from the combined system spectra.

However, current ground-based facilities are far from sufficient to provide the needed follow-up on the large number of exoplanets that  will be detected and characterized by these upcoming missions. The existing facilities either do not reach the performance needed for this follow-up or use 4- and 8-meters class telescopes. The latter are in high demand and expensive to operate, and the number of available observing time for exoplanet spectroscopic follow-up is simply too limited. This leaves a niche for new high-efficiency ground-based facilities, to allow those space missions to realize their scientific return at its full potential.

It is for this purpose that the MARVEL (Mercator Array for Radial VELocities) facility has been designed (see Raskin et al. 2020~\cite{MARVEL}; this conference). This paper is organized as follows. Section~\ref{sec:MARVEL} briefly presents the concept of MARVEL. In S
section~\ref{sec:sim}, we describe numerical simulations performed to investigate the optimization of the MARVEL facility. Our results are  presented in Section~\ref{sec:optiM} and our conclusions in  Section~\ref{sec:conc}

\section{MARVEL facility}\label{sec:MARVEL}

MARVEL is a high-precision radial-velocity spectroscopy facility using an array of four telescopes. The use of four telescopes provides more flexibility, by allowing to observe  either four different targets (one target per telescope) or to observe the same target with the four telescopes simultaneously, thus allowing the observation of fainter targets. MARVEL use 80-cm telescopes. Thanks to the high throughput possible with such small-size telescopes (Raskin et al. 2020~\cite{MARVEL}; this conference), MARVEL reaches an overall system efficiency comparable to that of 4m-class instruments+telescope system, for a much lower cost.

   \begin{figure} [ht]
   \begin{center}
   \begin{tabular}{c} 
   \includegraphics[height=8cm]{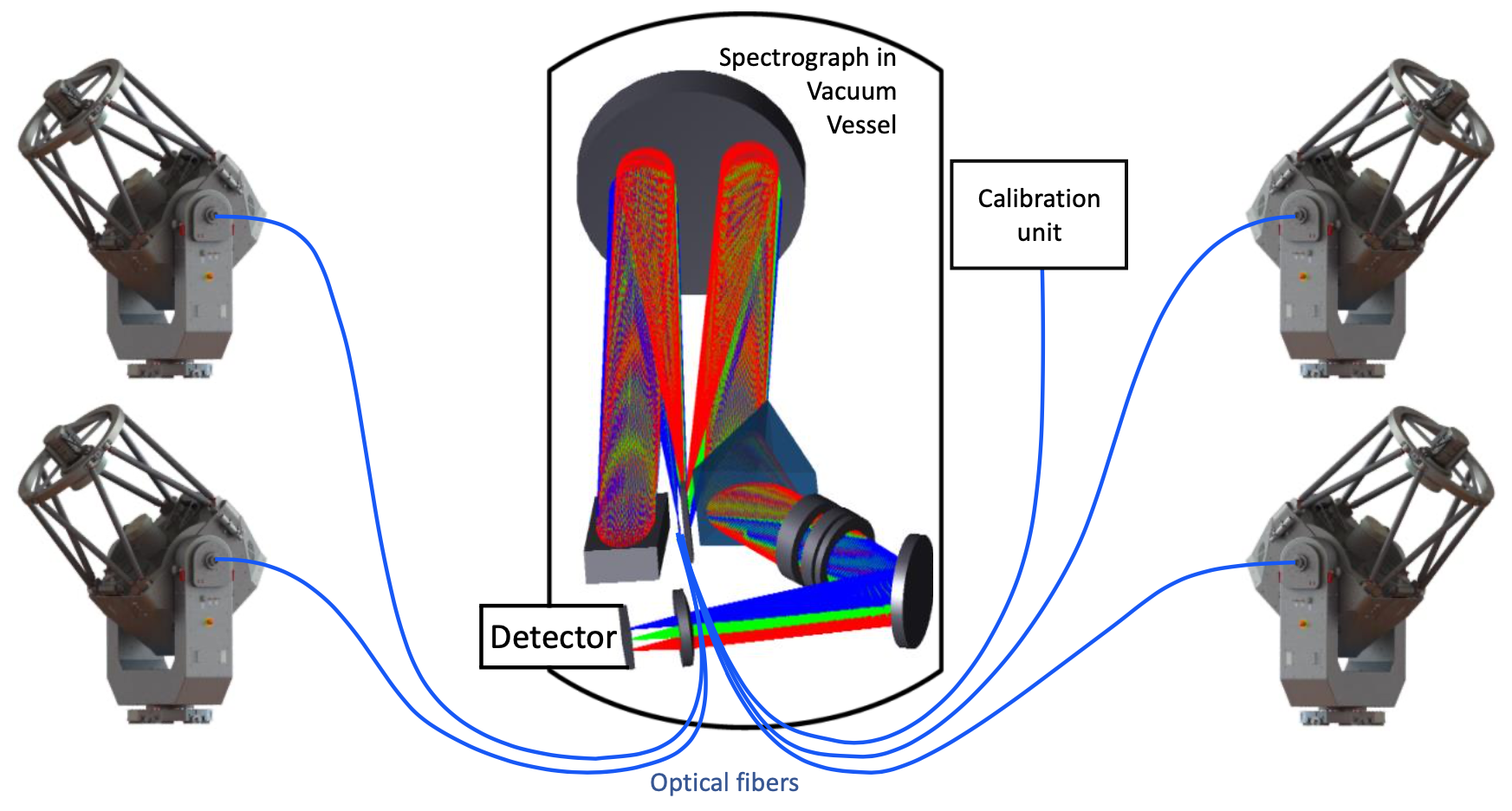}
   \end{tabular}
   \end{center}
   \caption[Conceptual layout of MARVEL.] 
   { Conceptual layout of MARVEL. \label{fig:scheme} }
   \end{figure} 

Figure~\ref{fig:scheme} displays the conceptual layout of MARVEL. Each of the four telescopes and the wavelength calibration unit is coupled to a common high-resolution échelle spectrograph by optical fibers. The spectrograph itself delivers a spectral resolution of at least $R = \lambda/\Delta\lambda = 90\,000$, and covers the spectral range from 390 to 920\,nm at once.

\begin{figure} [t!]
   \begin{center}
   \begin{tabular}{c} 
   \includegraphics[width=16cm]{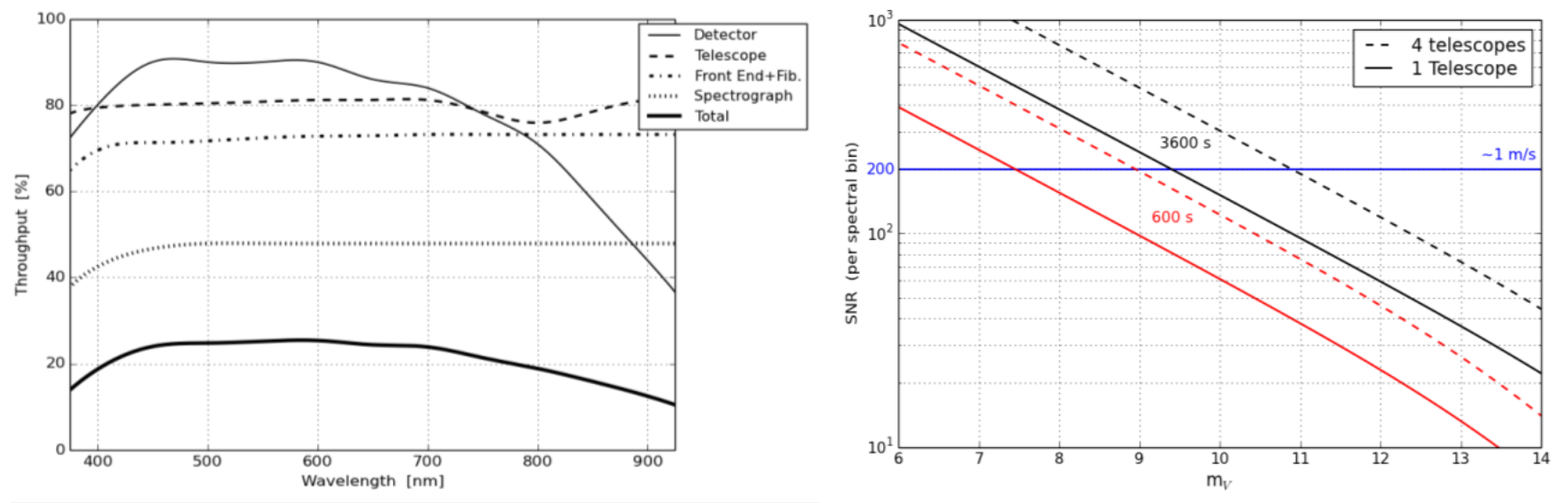}
   \end{tabular}
   \end{center}
   \caption[] 
   { Left: Throughput of the main components and of the total MARVEL instrument. Right: Estimated peak SNR per resolution element at 550 nm for various $V$-band magnitudes, for a single 80-cm telescope and for four such telescopes pointing to the same target. Exposure times are 10 minutes (red) and one hour (black). \label{fig:TandSNR} }
   \end{figure}

Thanks to a combination of already proven, state-of-the-art instrumental concepts, the total throughput of the facility is expected to be larger than 20\%, as shown in the left panel of Fig.~\ref{fig:TandSNR}.

Fig.~\ref{fig:TandSNR}-right shows the the SNR MARVEL will deliver as a function of the V-band stellar magnitude. The SNR of 200 (in blue) is the limit to suppress the photon noise to a level where we can reach a 1~m\,s$^{-1}$\, accuracy for an F-type star . As the figure shows, we can reach this accuracy down to a V-magnitude of 10.9, which is suitable for the follow-up of bright stars observed by the PLATO and ARIEL missions. More details on the design and performances can be found in Raskin et al. 2020~\cite{MARVEL}; this conference.
   
\section{Simulations}\label{sec:sim}

\subsection{Exoplanets simulation}\label{subsec:exosim}
In order to study and optimize the MARVEL operational model, we built a representative estimate of the exoplanet yield from PLATO, including the mass distribution of the candidates. To this aim, we used a typical stellar population expected in the Plato field. In this endeavor, we adopted the exoplanets occurrence of Dressing and Charbonneau (2015)~\cite{AFGK}, and we paired the host stars with exoplanets following the method described in Barclay et al. 2018~\cite{TESSyield}. To compute the RV signal induced by the exoplanets on the host stars, we adopted one typical mass for each bin of exoplanet radii in Dressing and Charbonneau (2015)~\cite{AFGK}, using the figure~3 of Chen and Kipping 2017~\cite{chen2017}. Table~\ref{tab:RtoM} provides the adopted typical mass for each bin.

\begin{table}[b]
\caption{Bins of exoplanets sizes and their adopted Representative masses} 
\label{tab:RtoM}
\begin{center}       
\begin{tabular}{|l|l|l|} 
\hline
\rule[-1ex]{0pt}{3.5ex}  Bin name & Radii range [$R_{\Earth}$] & Typical mass [$M_{\Earth}$]  \\
\hline
\rule[-1ex]{0pt}{3.5ex}  Giants & $6<R<22$ cm & 300   \\
\hline
\rule[-1ex]{0pt}{3.5ex}  Large Neptunes & $4<R<6$ & 30  \\
\hline
\rule[-1ex]{0pt}{3.5ex}  Small Neptunes & $2<R<4$ & 10  \\
\hline
\rule[-1ex]{0pt}{3.5ex}  Super-earths & $1.25<R<2$ & 3  \\
\hline 
\rule[-1ex]{0pt}{3.5ex}  Earth & $0.8<R<1.25$ & 1  \\
\hline 
\end{tabular}
\end{center}
\end{table}

From this simulated population of exoplanets, we used the PLATO redbook\footnote{\url{ https://sci.esa.int/s/8rPyPew }} and Marchiori et al. 2019\cite{PLATOdetect} to estimate the exoplanets that PLATO would detect.

Our simulations yielded an estimated  number of detections  similar but not identical to the total number of $\sim$ 4600 expected for the PLATO yield\footnote{\url{https://platomission.com/2018/06/04/planet-yield/}}. This can easily be explained by differences in the methodology and adopted assumptions. MARVEL being limited to follow-up Northern targets, we finally scaled our simulations to 2300 targets, detected by PLATO and accessible for spectroscopic follow-up from the ground by MARVEL.

\subsection{Integration time simulation}\label{subsec:intTsimu}

Coupling four telescopes to a single detector, the MARVEL facility offers various possible operational scenarios. It can be used in single-telescope mode (1-T), observing four different (brighter) stars simultaneously, or in four-telescope mode (4-T), observing the same (potentially fainter) target with the four telescopes simultaneously. To evaluate the relative importance of these two operational modes, we estimated the integration time necessary to achieve a high enough signal-to-noise ratio (SNR) to reach  a 1~m\,s$^{-1}$\, accuracy on the radial velocity. Three of the stellar parameters were kept free in the analysis: the effective temperature ($T_\mathrm{eff}$), the $V$-band magnitude ($V$), and the star's projected rotational velocity at the equator ($v_\mathrm{eq} \sin i$). The other atmospheric parameters were kept fixed at $\log g = 4.5$ (dwarfs), $v_\mathrm{micro} = 1$~km\,s$^{-1}$ and [Fe/H] = 0 (solar metallicity). The regions with strong telluric lines were excluded of the simulation.


   \begin{figure} [t!]
   \begin{center}
   \begin{tabular}{c} 
   \includegraphics[width=15cm]{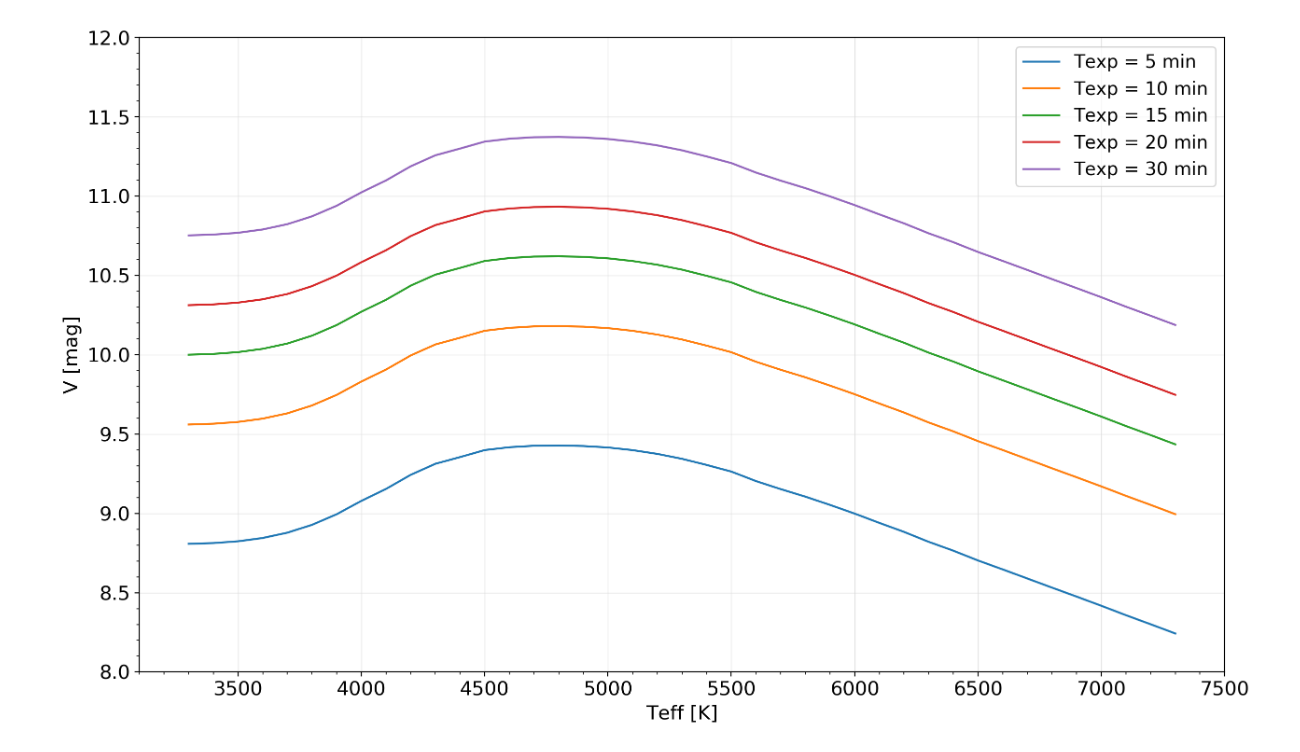}
   \end{tabular}
   \end{center}
   \caption[] 
   { Limiting Johnson $V$-band magnitudes to obtain a photon-noise-limited RV measurement accuracy of 1~m\,s$^{-1}$ for the  MARVEL  instrument  settings  (4-T  mode). The  parameters  of  the  synthetic  stellar  spectra  used  were $\log g = 4.5$, [Fe/H] = 0, $v_\mathrm{eq} \sin i$ = 2~km\,s$^{-1}$ and $v_\mathrm{micro} = 1$~km\,s$^{-1}$.  Wavelength regions with strong telluric lines were excluded. The different colors are for different exposure times (Texp; see legend). \label{fig:Tintvsmag}}
   \end{figure} 
   
Figure~\ref{fig:Tintvsmag} shows an example of the limiting $V$-magnitude allowing to reach a 1~m\,s$^{-1}$ accuracy as a function of the effective temperature for different integration times. 

\section{Optimization of MARVEL}\label{sec:optiM}

\subsection{The required RV accuracy}\label{subsec:RVaccuracy}

   \begin{figure} [t!]
   \begin{center}
   \begin{tabular}{c} 
   \includegraphics[width=12cm]{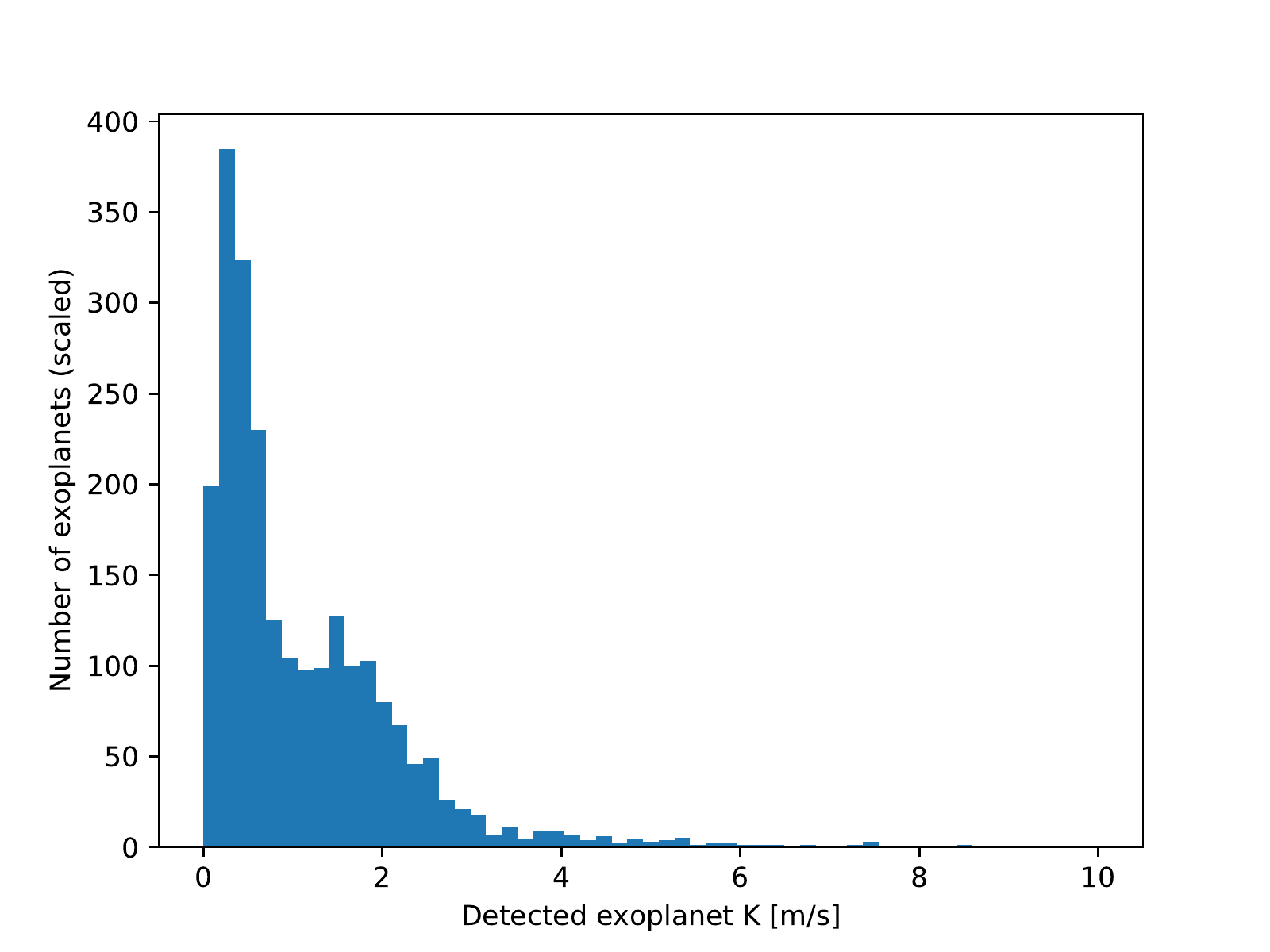}
   \end{tabular}
   \end{center}
   \caption[] 
   {Histogram of the semi-amplitude of the RV curve ($K$) of the host stars for a typical distribution of exo-solar systems that will be detected by PLATO. \label{fig:histRVd}}
   \end{figure}

For each of the simulated parameters of exoplanets that will be detected by PLATO (Sect.~\ref{subsec:exosim}), we computed the semi-amplitude ($K$) of the radial velocity curve of its host stars.
Figure~\ref{fig:histRVd} displays the distribution of $K$ in our simulations. It reveals that an RV measurement accuracy of 1~m\,s$^{-1}$\, will allow MARVEL to characterize a large part ($\sim$44\%) of the exoplanets detected with PLATO. If on the contrary MARVEL only reaches an accuracy of 2~m\,s$^{-1}$, its efficiency to perform PLATO follow-up would be highly degraded, with only $\sim$ 20\% of the exoplanets' host presenting an RV semi-amplitude larger than 2~m\,s$^{-1}$. To reach its purpose and to be able to lower the pressure on larger (4m+ class) facilities, it is therefore critical for MARVEL to reach this 1~m\,s$^{-1}$\, accuracy.

\subsection{The integration time: 1-T vs. 4-T}\label{subsec:1Tvs4T}

The multiple telescopes of MARVEL allow us some versatility with two observation modes. One mode uses the four telescopes on four different targets simultaneously, with one target per telescope (1-T mode). Another mode uses the four telescopes on a single target (4-T mode), allowing to collect four times more photons for the same integration time.

While the 1-T mode allows the observation of more targets at the same time, which is useful to reach the 12\,000 observations per year that MARVEL aims at, the 4T-mode would require four times less integration time on the same target to reach the suitable accuracy on the RV. Excluding any overheads, one would observe the same amount of targets in the same amount of time with both modes. But in order to average the effect of the stellar activity on the RV signal, one requires to observe for a minimum integration of time of about 20 minutes. So, on bright stars, one may reach the required SNR in less than 20 minutes with the 4-T mode, losing in efficiency in the observation planning, while with the 1-T mode, one would be able to efficiently plan our observation to avoid this unnecessary extra observing time.  Having to implement both observations mode will add complexity to the operational modes. So, to estimate the efficiency improvement that would bring the 1-T mode, we simulated the integration time required to reach the 1~m\,s$^{-1}$\, accuracy described in Sect.~\ref{subsec:intTsimu} for every exoplanet host-star, according to our simulations (sec.~\ref{subsec:exosim}).

\begin{figure} [t!]
   \begin{center}
   \begin{tabular}{c c} 
   \includegraphics[width=7.9cm]{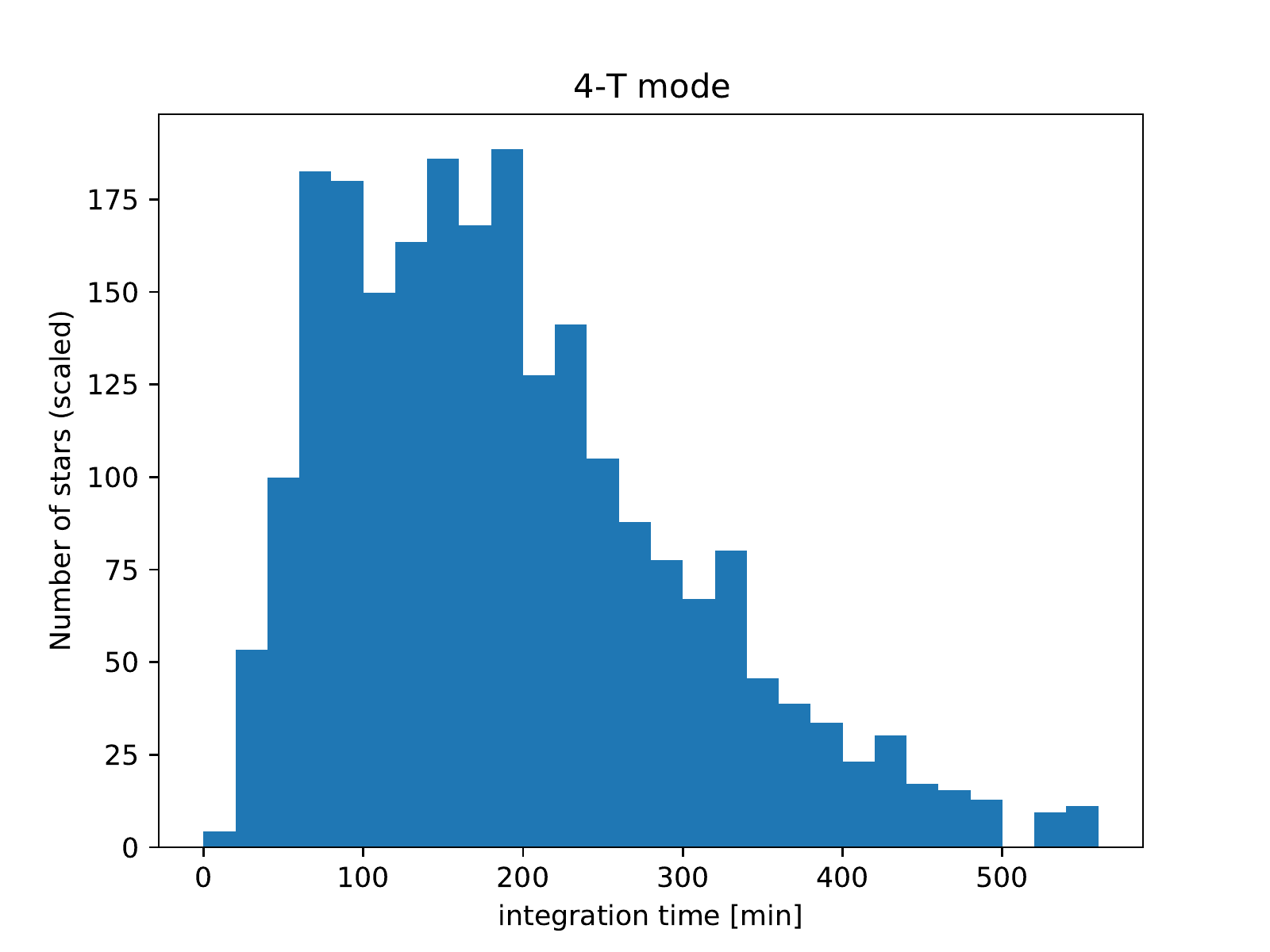}
   &
   \includegraphics[width=7.9cm]{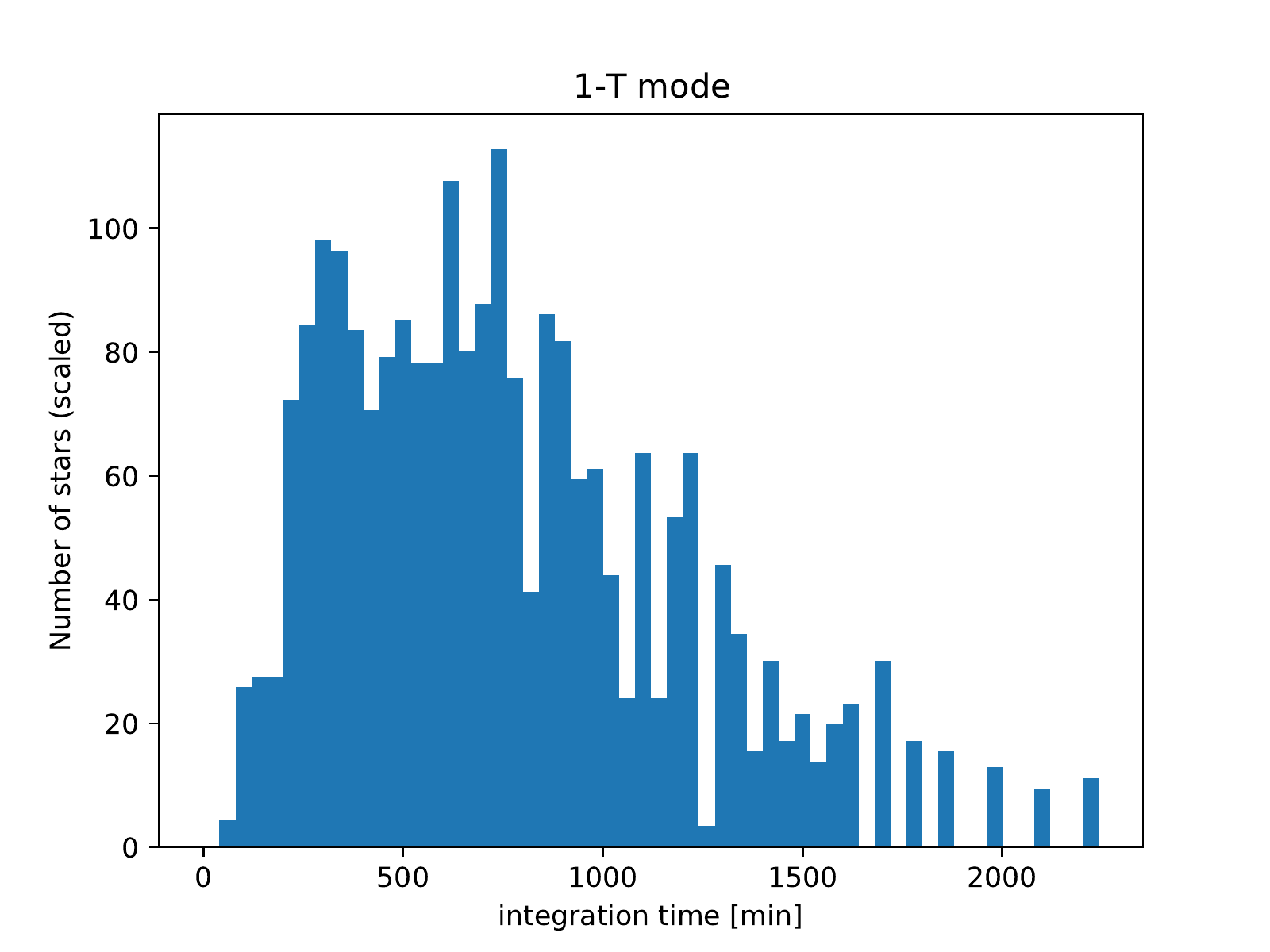}
   \end{tabular}
   \end{center}
   \caption[] 
   {Histograms of the required integration times needed to reach a 1~m\,s$^{-1}$\, RV accuracy. \textbf{Left:} 4-T mode. \textbf{Right:} 1-T mode \label{fig:histTint}}
   \end{figure}

Figure~\ref{fig:histTint} displays the histogram of the required integration time needed to reach 1~m\,s$^{-1}$\, on  stars hosting an exoplanet that would have been detected by PLATO. 

In order to measure the efficiency brought by the 1-T mode, we need to look at the number of stars in the first 20 minutes bin of the 4T-mode. Only a few (4) stars would require 20 minutes or less to reach the 1~m\,s$^{-1}$\, accuracy on the RV. So, the 1-T mode would be more efficient for those 4 stars, requiring less total observing time to observe those for stars. The efficiency of the implementation of the 1-T mode is however negligible, relative to the total number of stars.

From this study, we show that the implementation of the 1-T mode is not required. We will discuss the utility of the 1-T mode implementation furthermore in Sect.~\ref{sec:conc}.

\section{Discussion and Conclusions}\label{sec:conc}

The results of Sect.~\ref{subsec:RVaccuracy} show that to relieve the large facility of a large part of the PLATO follow-up, MARVEL needs to reach an accuracy of 1~m\,s$^{-1}$. This accuracy is fortunately achievable thanks to the instrument design presented in Raskin et al. 2020~\cite{MARVEL} (this conference). The present result is based on a simplified simulation of the number and mass distribution of the exoplanets that PLATO should detect. The present results can be straightforwardly updated once the PLATO input catalog and/or the PLATO detection yield simulation will be made public. In addition, the use of the only five typical masses to compute our RV signal could be improved, using a linear relation between radius and mass for each of the five bins we used. Yet, we do not expect that these improvements will significantly change the qualitative results of the present simulations.

Section~\ref{subsec:1Tvs4T} shows that, on the one hand, the gain in efficiency brought by the implementation of a versatile 1-T mode is negligible when aiming at a 1~m\,s$^{-1}$\, accuracy on each exoplanet hosting star detected by PLATO. On the other, the implementation of  the observing mode would add unnecessary complexity to the operational model. However, MARVEL does not have to reach the 1~m\,s$^{-1}$\, accuracy on every target. For instance, one would require much less precise measurements to characterize a hot Jupiter. But efficiently use the 1-T mode for those cases would require a prior knowledge of the RV accuracy one needs to reach for each target. A further study is required to determined if the portion of targets that required less than 20 minutes of observing time with 4-T mode to reach the accuracy on the RV at a level of the RV induced by the exoplanet justify the implementation of the 1-T mode.

\section*{Acknowledgment}

The MARVEL project and team acknowledge support from the Fund for Scientific Research of Flanders (FWO) under the  Large Infrastructure Program grant I011020N.

\appendix    


\bibliography{report} 
\bibliographystyle{spiebib} 

\end{document}